\documentclass[10pt,conference]{IEEEtran}

\usepackage[T1]{fontenc}
\usepackage{booktabs}
\usepackage{cite}
\usepackage[font=footnotesize]{caption}
\usepackage{comment}
\usepackage{enumitem}
\usepackage{hyperref}
\usepackage{inconsolata}
\usepackage[frozencache]{minted}
\usepackage{multirow}
\usepackage{pifont}
\usepackage{rotating}
\usepackage{textgreek}
\usepackage{tikz}
\usepackage{xurl}  %

\usetikzlibrary{arrows,decorations.pathreplacing,tikzmark}
\tikzset{>=stealth}

\setminted{
  autogobble,
  breaklines,
  fontsize  = \scriptsize,
  frame     = single,
  linenos,
  numbersep = 0.5ex,
}

\definecolor{peekcolor}{rgb}{0.760, 0.890, 1}
\definecolor{readcolor}{gray}{0.888}
\newcommand{\colorpeek}[1]{{\setlength{\fboxsep}{1pt}\colorbox{peekcolor}{#1}}}
\newcommand{\colorread}[1]{{\setlength{\fboxsep}{1pt}\colorbox{readcolor}{#1}}}

\newcommand{\mycircled}[1]{\raisebox{-1pt}{\ding{\numexpr171+#1\relax}}}  %

\hypersetup{
  colorlinks = true,  %
  urlcolor   = black, %
  linkcolor  = black, %
  citecolor  = black  %
}

\newcommand\kernellocreduced{22}

\newcommand\hostlocreduced{51}

\newcommand\csimtimespeedup{3.2}

\newcommand\hlstimespeedup{6.8}

\begin{document}

\setlength{\abovedisplayskip}{1em}
\setlength{\belowdisplayskip}{1em}
\setlength{\textfloatsep}{1em}
\renewcommand{\baselinestretch}{0.98}

\title{Extending High-Level Synthesis for Task-Parallel Programs}

\author{
  \IEEEauthorblockN{
    Yuze Chi\IEEEauthorrefmark{1},
    Licheng Guo\IEEEauthorrefmark{1},
    Jason Lau\IEEEauthorrefmark{1},
    Young-kyu Choi\IEEEauthorrefmark{1}\IEEEauthorrefmark{2},
    Jie Wang\IEEEauthorrefmark{1},
    Jason Cong\IEEEauthorrefmark{1}
  }
  \IEEEauthorblockA{
    \IEEEauthorrefmark{1}University of California, \textit{Los Angeles},
    \IEEEauthorrefmark{2}Inha University\\
    \texttt{\{chiyuze,cong\}@cs.ucla.edu}
  }
}

\maketitle

\thispagestyle{plain}
\pagestyle{plain}

\begin{abstract}
  C/C++/OpenCL-based high-level synthesis (HLS) becomes more and more popular for
  field-programmable gate array (FPGA) accelerators in many application
  domains in recent years,
  thanks to its competitive quality of results (QoR) and short development cycles
  compared with the traditional register-transfer level design approach.
Yet, limited by the sequential C semantics,
  it remains challenging to adopt the same highly productive high-level
  programming approach in many other application domains,
  where coarse-grained tasks run in parallel and communicate with each other at
  a fine-grained level.
While current HLS tools do support task-parallel programs,
  the productivity is greatly limited
  \mycircled{1}~in the code development cycle due to
  the poor programmability,
  \mycircled{2}~in the correctness verification cycle due to
  restricted software simulation,
  and \mycircled{3}~in the QoR tuning cycle due to slow code generation.
Such limited productivity often defeats the purpose of HLS and hinder
  programmers from adopting HLS for task-parallel FPGA accelerators.

In this paper,
  we extend the HLS C++ language and present a fully automated framework with
  programmer-friendly interfaces,
  unconstrained software simulation,
  and fast hierarchical code generation to overcome these limitations and
  demonstrate how task-parallel programs can be productively supported in HLS.
Experimental results based on a wide range of real-world task-parallel programs
  show that, on average,
  the lines of kernel and host code are reduced by \kernellocreduced\% and
  \hostlocreduced\%,
  respectively, which considerably improves the programmability.
The correctness verification and the iterative QoR tuning cycles
  are both greatly shortened by \csimtimespeedup{\texttimes} and
  \hlstimespeedup{\texttimes}, respectively.
Our work is open-source at \texttt{\url{https://github.com/UCLA-VAST/tapa/}}.

\end{abstract}

\section{Introduction}

C/C++/OpenCL-based high-level synthesis (HLS)~\cite{tcad11-hls} has been adopted
  rapidly by both the academia and the industry for programming
  field-programmable gate array (FPGA) accelerator design in many application
  domains, e.g.,
  machine learning~\cite{dac16-dnnfreq, iccad18-polysa, fpga19-heterocl},
  scientific computing~\cite{fpga18-stencil-blocking, taco19-dcmi,
  dac20-soda-cr, cgo21-stencilflow},
  and image processing~\cite{iccad18-soda, taco17-halide-hls,
  fpga20-heterohalide}.
Compared with the traditional register-transfer level (RTL)
  paradigm where the debug turnaround time of even simple
  applications~\cite{tapa-apps} can take tens of minutes,
  with HLS, programmers can follow a rapid development cycle.
Programmers can write code in C
  and leverage fast software simulation to verify the functional correctness.
The debug turnaround time for such a correctness verification cycle can take as
  few as just one second instead of tens of minutes,
  allowing functionalities to be iterated at a fast pace.
Once the HLS code is functionally correct,
  programmers can then generate RTL code,
  evaluate the quality of results (QoR) based on the generated performance and
  resource reports,
  and modify the HLS code accordingly.
Such a QoR tuning cycle typically takes only a few minutes for a simple design
  or a component in a modular design.

Thanks to the advances in HLS scheduling
  algorithms~\cite{dac06-sdc, fpga19-easy,
  fpga20-dyn-sta-hls, dac19-thread-weaving, mlsys20-autophase}
  and timing
  optimizations~\cite{fpt19-reg-hls, dac20-hls-timing, fpga20-buffer-dataflow,
  fpga21-autobridge},
  HLS can not only shorten the development cycle,
  but also generate programs that are often competitive in cycle
  count~\cite{dac18-autoaccel},
  and more recently in clock frequency as
  well~\cite{dac20-hls-timing, fpga21-autobridge}.
Moreover,
  FPGA vendors provide host drivers and communication interfaces for kernels
  designed in HLS~\cite{man-vivado-hls,man-intel-opencl},
  further reducing programmers' burden to integrate and offload workload to FPGA
  accelerators.

However, not all programs are created equal for HLS.
Data-parallel programs can be easily programmed following the sequential C
  semantics,
  which enables such applications to be quickly designed and iterated in the
  fast correctness verification cycle and QoR tuning cycle.
In contrast,
  task-parallel programs are not supported by the native C semantics,
  and the productivity provided by current HLS tools is greatly limited
  for the following reasons:
\begin{itemize}[topsep=0pt,leftmargin=*]

\item \textit{Poor programmability}.
Due to the lack of convenient application programming interfaces (API),
  programmers are often forced to write more code than necessary.
For example,
  for an accelerator with PEs connected through a simple on-chip network,
  a network node needs to forward packets based on their content (header) and
  the availability of output ports.
Without an API to read packets without consuming
  them (i.e., ``\textit{peek}'') from the ports,
  programmers have to manually and carefully create a buffer and maintain a
  small state machine to keep track of incoming packets.
This not only elongates the development cycle,
  but also is error-prone.

\item \textit{Restricted software simulation}.
As the key to fast correctness verification,
  software simulation is not always available to task-parallel programs.
For example,
  Vivado HLS software simulation does not support Cannon's
  algorithm~\cite{ics97-cannon} because its sequential execution of tasks cannot
  correctly simulate feedback loops in data paths,
  while Intel OpenCL simulator does not support more than 256 concurrent
  kernels~\cite{man-intel-opencl}.
Unavailability of software simulation forces programmers to resort to RTL
  simulation for correctness verification,
  significantly elongating the development cycle.

\item \textit{Slow code generation}.
We found that current HLS compilers do not support hierarchical code generation
  for task-parallel programs.
Instead,
  they treat all tasks as a monolithic design and process each instance of the
  same task as if they were different.
For designs that instantiate the same task many times
  (e.g., in a systolic array),
  this leads to repetitive compilation on each task and unnecessarily slows down
  code generation.
Programmers can manually synthesize tasks separately and
  instantiate them in RTL,
  but doing so requires debugging RTL code,
  which is time-consuming and error-prone.
We think such processes should be automated.

\end{itemize}

Limited productivity support for task-parallel programs significantly elongates the
  development cycles and undermines the benefits brought by HLS.
One may argue that programmers should always go for data-parallel
  implementations when designing FPGA accelerators using HLS,
  but data-parallelism may be inherently limited, for example,
  in applications involving graphs.
Moreover,
  researches show that even for data-parallel applications like neural
  networks~\cite{iccad18-polysa} and stencil computation~\cite{iccad18-soda},
  task-parallel implementations show better scalability and higher frequency
  than their data-parallel counterparts
  due to the localized communication pattern~\cite{fccm18-latte}.
In fact,
  at least 6 papers~\cite{fpga20-heterohalide,fpga20-cancer,fpga20-lstm,
  fpga20-dnn,fpga20-gemm,fpga20-boyi} among the 28 research papers published in
  the ACM FPGA 2020 conference use task-parallel implementation with HLS,
  and another 3 papers~\cite{fpga20-graphact,fpga20-switch,fpga20-samplesort}
  use RTL implementation that would have required task-parallel implementation
  if written in HLS.

In this paper,
  we extend the HLS C++ language and present our framework,
  TAPA (\textbf{ta}sk-\textbf{pa}rallel)\footnote{
    While a prior work TAPAS~\cite{micro18-tapas} and our work TAPA share
      similarity in name,
      our work focuses on statically mapping tasks to hardware,
      yet TAPAS specializes in dynamically scheduling tasks.
  },
  as a solution to the aforementioned limitations of HLS productivity.
Our contributions include:

\begin{itemize}[leftmargin=*]

\item \textbf{Convenient programming interfaces}:
We show that,
  with peeking and transactions added to the programming interfaces,
  TAPA can be used to program task-parallel kernels with \kernellocreduced\%
  reduction in lines of code (LoC) on average.
By unifying the interface used for the kernel and host,
  TAPA further reduces the LoC on the host side by \hostlocreduced\% on average.

\item \textbf{Unconstrained software simulation}:
We demonstrate that our proposed simulator can correctly simulate task-parallel
  programs that existing software simulators fail to simulate.
Moreover,
  the correctness verification cycle can be shortened by a factor of
  \csimtimespeedup{\texttimes} on average.

\item \textbf{Hierarchical code generation}:
We show that by modularizing a task-parallel program and using a hierarchical
  approach,
  RTL code generation can be accelerated by a factor of
  \hlstimespeedup{\texttimes} on our
  server with 32 hyper-threads.

\item \textbf{Fully automated open-source framework}: TAPA is open-source at
  \texttt{\url{https://github.com/UCLA-VAST/tapa/}}.
\end{itemize}

Table~\ref{tab:related-work} summarizes the related work.
Among all general HLS tools (Section~\ref{sec:related-work-hls}) and
  streaming frameworks (Section~\ref{sec:related-work-streaming}):
\mycircled{1}~None of them supports peeking in their kernel APIs;
  only Intel HLS \texttt{stream} and Vivado HLS \texttt{axis} support
  transactions;
  only Merlin allows the accelerator kernel to be called from the host as if it
  is a C/C++ function.
\mycircled{2}~Vivado HLS, Merlin, and both streaming frameworks
  (ST-Accel~\cite{fccm18-staccel} and Fleet~\cite{asplos20-fleet}) execute tasks
  sequentially for simulation, which works on limited applications,
  while others launch one thread per task instance,
  which does not scale well.
\mycircled{3}~All general HLS tools treat a task-parallel program as a
  monolithic design and generate RTL code for each instance of task separately,
  except that Vivado HLS \texttt{axis} allows programmers to manually
  instantiate tasks using a configuration file when running logic synthesis and
  implementation.
To the best of our knowledge,
  TAPA is the only work that provides convenient programming interfaces,
  unconstrained software simulation,
  and hierarchical code generation for general task-parallel programs on FPGAs
  using HLS.

\begin{table}[!t]
  \centering
  \caption{Summary of related work.}
  \label{tab:related-work}
  \resizebox{\linewidth}{!}{
    \begin{tabular}{@{}l@{}p{1.3em}p{1.6em}p{2.6em}l@{}l@{}}
      \toprule
      \multirow{3}{*}{Related Work}                   & \multicolumn{3}{c}{Programmability}   & \multirow{3}{*}{\shortstack[l]{Software\\Simulation}} & \multirow{3}{*}{\shortstack[l]{RTL Code\\Generation}} \\
                                                      & Peek-ing & Trans-action & Host Iface. &                                                       &                                                       \\
      \midrule
      Fleet~\cite{asplos20-fleet}                     & No       & No           & N/A         & Sequential                                            & N/A                                                   \\
      Intel HLS (\texttt{pipe})                       & No       & No           & N/A         & Multi-thread~~                                        & Monolithic                                            \\
      Intel HLS (\texttt{stream})                     & No       & Yes          & N/A         & Multi-thread                                          & Monolithic                                            \\
      Intel OpenCL                                    & No       & No           & OpenCL      & Multi-thread                                          & Monolithic                                            \\
      LegUp~\cite{fpga11-legup,tvlsi17-legup-pthread} & No       & No           & N/A         & Multi-thread                                          & Monolithic                                            \\
      Merlin~\cite{islped16-merlin}                   & No       & No           & C++         & Sequential                                            & Monolithic                                            \\
      ST-Accel~\cite{fccm18-staccel}                  & No       & No           & VFS         & Sequential                                            & Hierarchical                                          \\
      Vivado HLS (\texttt{ap\_fifo})~~                & No       & No           & OpenCL      & Sequential                                            & Monolithic                                            \\
      Vivado HLS (\texttt{axis})                      & No       & Yes          & OpenCL      & Multi-thread                                          & Manual                                                \\
      Xilinx OpenCL                                   & No       & No           & OpenCL      & Multi-thread                                          & Monolithic                                            \\
      \midrule
      TAPA                                            & Yes      & Yes          & C++         & Coroutine                                             & Hierarchical                                          \\
      \bottomrule
    \end{tabular}
  }  %
\end{table}

\section{Background}

\subsection{Task-Parallel Program}
\label{sec:task-parallelism}

Task-level parallelism is a form of parallelization of computer programs across
  multiple processors.
In contrast to data parallelism where the workload is partitioned on data
  and each processor executes the same program
  (e.g., OpenMP~\cite{cse98-openmp}),
  different processors in a task-parallel program often behave differently,
  while data are passed between processors.
Examples of task-parallel programs include image
  processing pipelines~\cite{taco17-halide-hls, iccad18-soda,
  fpga20-heterohalide},
  graph processing~\cite{fpga16-fpgp, fpga17-foregraph, tpds19-hitgraph,
  fccm19-worklist-graph},
  and network switching~\cite{fpga20-switch}.
Task-parallel programs are often described using dataflow
  models~\cite{cacm78-csp, ifip74-kpn, ieee87-sdf, thesis93-boolean-dataflow,
    acmcs77-petrinet},
  where tasks are called \textit{processes}.
Processes communicate only through unidirectional \textit{channels}.
Data exchanged between channels are called \textit{tokens}.
In this paper,
  we borrow the terms \textit{channel} and \textit{token},
  and focus on the problem of statically mapping tasks to hardware.
That is,
  instances of tasks are synthesized to different areas in an FPGA accelerator.
We plan to address dynamic
  scheduling~\cite{tvlsi17-legup-pthread,micro18-tapas,asplos20-chronos}
  in our future work.

\subsection{A Motivating Example}
\label{sec:motivating-example}
An on-chip ring network is a commonly used topology to provide
  all-to-all interconnection among many task-parallel processing elements (PE)
  in a single FPGA accelerator,
  which is particularly useful in graph processing~\cite{iclr17-gcn,
    iclr20-graphzoom, tr98-pagerank, im09-community, nips12-ego,
    icde16-nxgraph, tcad18-graphh} where each
  vertex may be connected to any other vertices.
A ring network has the advantages of simplicity and high routability,
  but implementing a customized ring network in HLS faces several issues that
  make such designs verbose to write, hard to read, and error-prone.
In this section,
  we use a simplified real-world design to illustrate the productivity issues
  for implementing such a ring network in HLS,
  which serves as a motivating example for our work.

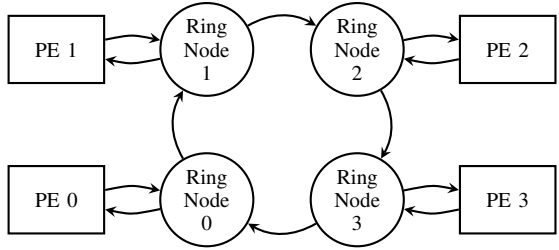
\begin{figure}[!ht]
  \begin{tikzpicture}[thick,font=\footnotesize]
    \path
      (0, -2) node [draw,inner sep=1em] (pe0) {PE 0}
      (0,  0) node [draw,inner sep=1em] (pe1) {PE 1}
      (6,  0) node [draw,inner sep=1em] (pe2) {PE 2}
      (6, -2) node [draw,inner sep=1em] (pe3) {PE 3}
      (2, -2) node [align=center,circle,draw,inner sep=0.5ex] (node0) {Ring\\Node\\0}
      (2,  0) node [align=center,circle,draw,inner sep=0.5ex] (node1) {Ring\\Node\\1}
      (4,  0) node [align=center,circle,draw,inner sep=0.5ex] (node2) {Ring\\Node\\2}
      (4, -2) node [align=center,circle,draw,inner sep=0.5ex] (node3) {Ring\\Node\\3}
    ;

    \draw [->] (pe0) .. controls (1, -1.8) .. (node0);
    \draw [->] (node0) .. controls (1, -2.2) .. (pe0);
    \draw [->] (pe1) .. controls (1, 0.2) .. (node1);
    \draw [->] (node1) .. controls (1, -0.2) .. (pe1);
    \draw [->] (node2) .. controls (5, 0.2) .. (pe2);
    \draw [->] (pe2) .. controls (5, -0.2) .. (node2);
    \draw [->] (node3) .. controls (5, -1.8) .. (pe3);
    \draw [->] (pe3) .. controls (5, -2.2) .. (node3);

    \draw [bend left,->] (node0) to (node1);
    \draw [bend left,->] (node1) to (node2);
    \draw [bend left,->] (node2) to (node3);
    \draw [bend left,->] (node3) to (node0);
  \end{tikzpicture}
  \centering
  \caption{An accelerator with 4 PEs connected via a ring network.}
  \label{fig:ring-network}
\end{figure}

Fig.~\ref{fig:ring-network} shows an example where PEs in an accelerator are
  interconnected via a ring network.
In this example, network nodes form a cyclic ring,
  and each ring node is connected to a PE via a bidirectional link.
Each PE can send packets to other PEs through its associated node,
  specifying its destination PE in the packet header.
Each node forwards packets either to its next node or to its associated PE,
  based on the packet header.
We assume packets are sent infrequently and channels between nodes are
  provisioned so that they will never be full.
Furthermore,
  we would like to insert packets from PEs to the network ASAP
  so that PEs will not stall due to back pressure from the ring nodes.
While such a ring node can be written using Vivado HLS
  (Listing~\ref{lst:ring-hls-comm}),
  we found that the followings are missing or hard-to-use in the HLS tools and
  significantly degrade the productivity.

\subsubsection{Peeking}
Peeking is defined as reading a token from a channel without consuming it.
Compared with the normal destructive read,
  peeking is non-destructive because the token may be read many times.
For example, in our ring network,
  when Node 1 receives incoming packets from both PE 1 (via \texttt{pe\_in}) and
  Node 0 (via \texttt{node\_in}),
  it will forward the packet from PE 1 to Node 2 (via \texttt{node\_out}) to
  prevent PE 1 from being stalled due to back pressure.
In the same clock cycle,
  the packet from Node 0 cannot be forwarded unless the destination of that
  packet is PE 1 (via \texttt{pe\_out}),
  because we cannot write two tokens to the same output channel
  (\texttt{node\_out}) in the same clock cycle.
This requires us to conditionally read tokens based on the content of tokens.
Without a \texttt{peek} API,
  one has to manually maintain a buffer for the incoming values,
  as shown in Line~\ref{lln:peek_top}--\ref{lln:peek_bottom} of
  Listing~\ref{lst:ring-hls-comm}.
This not only increases the programming burden,
  but also makes the design prone to errors in state transitions of the buffer.

\subsubsection{Transactions}
A sequence of tokens may constitute a single logical communication transaction.
Using the same ring network example,
  we consider the whole accelerator execution as a logical communication
  transaction,
  and let each PE control the termination of each \texttt{RingNode},
  as shown in Line~\ref{lln:eot} of Listing~\ref{lst:ring-hls-comm}.
Without an \texttt{eot} API,
  one has to manually add a special bit to the data structure to indicate the
  \textit{end-of-transaction} (Line~\ref{lln:eot_top}--\ref{lln:eot_bottom} of
  Listing~\ref{lst:ring-hls-comm}).
Note that the \texttt{Pkt} struct may be used elsewhere,
  thus it may be infeasible to add the \texttt{eot} bit directly to the
  \texttt{Pkt} struct.
Moreover,
  determining the end of transaction must be a peek operation;
  otherwise,
  the HLS compiler will be unable to schedule the exit condition in the first
  stage of pipeline,
  leading to II greater than 1.
This further complicates the HLS implementation
  (Listing~\ref{lst:ring-hls-comm}).

\subsubsection{System integration}
To offload computation kernel from the host CPU to PCIe-based FPGA accelerators,
  programmers need to write host-side code to interface the accelerator kernel
  with the host.
FPGA vendors adopt the OpenCL standard to provide such a functionality.
While the standard OpenCL host-kernel interface infrastructure relieves
  programmers from writing their own operating system drivers and low-level
  libraries,
  it is still inconvenient and hard-to-use.
Programmers often have to write and debug tens of lines of code just to set up
  the host-kernel interface.
This includes manually setting up environmental variables for simulation,
  creating, and maintaining OpenCL \texttt{Context}, \texttt{CommandQueue},
  \texttt{Program}, \texttt{Kernel},
  etc.\ data structures~\cite{vitis-opencl-host}.
Task-parallel accelerators often make the situation worse because the parallel
  tasks are often described as distinct OpenCL kernels~\cite{man-intel-opencl},
  which significantly increases the programmers' burden on managing multiple
  kernels in the host-kernel interface.
In our experiments, more than 60 lines of host code are created just
  for the host-kernel integration, which constitute more than 20 percent of the
  whole source code.
Yet, what we want is just a single function invocation of the
  synthesized FPGA bitstream given proper arguments.

\subsubsection{Software simulation}
C does not have explicit parallel semantics by itself.
Vivado HLS uses the dataflow model and allows programmers to instantiate tasks
  by invoking each of them sequentially~\cite{man-vivado-hls}.
While this is very concise to write (Listing~\ref{lst:ring-hls-inst}),
  it leads to incorrect simulation results because the communication
  between a ring node and its corresponding PE is bidirectional,
  yet sequential execution can only send tokens from nodes to PEs because of
  their invocation order.
This problem was also pointed out in~\cite{fpga19-flash}.
In order to run software simulation correctly,
  the programmer can change the source code to run tasks in multiple threads,
  but doing so requires the same piece of task instantiation code to be
  written twice for synthesis and simulation,
  reducing productivity.
While there exist other tools (e.g.~\cite{man-intel-opencl}) that can run tasks
  in parallel threads and do not have the same correctness problem,
  we will show in Section~\ref{sec:exp-simulation} that such simulators do not
  scale well when the number of task instances increases.

\subsubsection{RTL code generation}
In our ring network example, the same ring node is instantiated many times.
While state-of-the-art HLS compilers can recognize multiple instances of the
  same function and reuse HLS results for regular non-task-parallel programs,
  task-parallel programs are always treated as a monolithic one.
This means instances of the same task in a task-parallel program are treated as
  if they were different,
  possibly in order to explore different communication interfaces of each
  instance.
This significantly elongates the code generation time when the number of
  instances is large (Section~\ref{sec:exp-codegen}).
We can manually do hierarchical code generation,
  i.e., synthesize each task separately and connect the generated RTL code,
  but doing so forces us to debug RTL code and spend tens of minutes to verify
  the correctness for each code modification,
  thus defeats the purpose for adopting HLS.

In this paper,
  we present the TAPA framework and address these challenges by providing
  convenient programming interfaces, unconstrained software simulation,
  and hierarchical code generation.

\section{TAPA Programming Model and Interfaces}

\subsection{Hierarchical Programming Model}
\label{sec:dfsm-model}

TAPA uses a hierarchical programming model.
Each task is either a leaf that does not instantiate any channels or tasks,
  or a collection of tasks and channels with which the tasks communicate.
A task that instantiates a set of tasks and channels is called the
  \textit{parent task} for that set.
Correspondingly,
  the instantiated tasks are the \textit{children tasks} of their parent,
  which may be parents of their own children.
Each channel must be connected to exactly two tasks.
One of the tasks must act as a \textit{producer} and the other must act as a
  \textit{consumer}.
The producer \textit{streams} tokens to the consumer via the channel
  in the first-in-first-out (FIFO) order.
Each task is implemented as a C++ function,
  which can communicate with each other via the
  \textit{communication interface}.
A parent task instantiates channels and tasks using the
  \textit{instantiation interface},
  and waits until all its children tasks finish.
One of the tasks is designated as the \textit{top-level task},
  which defines the communication interfaces external to the FPGA accelerator,
  i.e., the \textit{system integration interface}.
\begin{tikzpicture}[font=\small,red,thick,overlay,remember picture]
  \node (a) at (pic cs:eot_top) {};
  \draw [decorate,decoration=brace] (a.north) -- (pic cs:eot_bottom)
    node [align=left,midway,right,xshift=1ex] {
      Auxiliary \texttt{struct} for termination control;\\
      \texttt{eot} stands for ``end of transaction''.
    };

  \node (a) at (pic cs:buf_top) {};
  \draw [decorate,decoration=brace] (a.north) -- (pic cs:buf_bottom)
    node [align=left,midway,right,xshift=1ex] {
      Manually maintained input\\
      buffers to implement non-\\
      destructive read (i.e., peek).
    };

  \node (a) at (pic cs:state_top) {};
  \draw [decorate,decoration=brace] (a.north) -- (pic cs:state_bottom)
    node [align=left,midway,right,xshift=1ex] {
      Manually\\update\\buffers.
    };
\end{tikzpicture}

\begin{listing}[!t]
  \begin{minted}[escapeinside=||,texcomments]{cpp}
    struct PktEoT { |\tikzmark{eot_top}\label{lln:eot_top}|
      Pkt pkt;
      bool eot;
    };              |\tikzmark{eot_bottom}\label{lln:eot_bottom}|
    void RingNode(stream<Pkt>& node_in,  stream<PktEoT>& pe_in,
                  stream<Pkt>& node_out, stream<Pkt>& pe_out) {
      Pkt node_pkt;                |\tikzmark{buf_top}\label{lln:peek_top}|
      bool node_pkt_valid = false;
      PktEoT pe_pkt;
      bool pe_pkt_valid = false;   |\tikzmark{buf_bottom}|
      while (!(|\colorpeek{pe\_pkt\_valid && pe\_pkt.eot}\label{lln:eot}|)) {
        if (!pe_pkt_valid)                            |\tikzmark{state_top}|
          pe_pkt_valid = pe_in.read_nb(pe_pkt);
        if (!node_pkt_valid)
          node_pkt_valid = node_in.read_nb(node_pkt); |\tikzmark{state_bottom}\label{lln:peek_bottom}|
        if (pe_pkt_valid) {
          node_out.write(|\colorread{pe\_pkt}|.pkt);
          |\colorread{pe\_pkt\_valid = false;}|
          if (node_pkt_valid && IsForThisNode(|\colorpeek{node\_pkt}|)) {
            pe_out.write(|\colorread{node\_pkt}|);
            |\colorread{node\_pkt\_valid = false;}|
          }
        } else if (node_pkt_valid) {
          Pkt pkt = |\colorread{node\_pkt}|;
          |\colorread{node\_pkt\_valid = false;}|
          (IsForThisNode(pkt) ? pe_out : node_out).write(pkt);
        }
      }  // Highlighted are \colorread{destructive read operations} and
    }    // \colorpeek{non-destructive read (peek) operations}.
  \end{minted}
  \caption{Ring network node written in Vivado HLS.}
  \label{lst:ring-hls-comm}
\end{listing}

\begin{listing}[!t]
  \begin{minted}[escapeinside=||]{cpp}
    void Kernel(...) {
      stream<Pkt, 2> node_0_1, node_1_2, ...
      stream<Pkt, 2> from_pe_0, to_pe_0, from_pe_1, to_pe_1, ...
      // Instantiates other channels...
    #pragma HLS dataflow
      RingNode(node_0_1, node_1_2, from_pe_1, to_pe_1);
      RingNode(node_1_2, node_2_3, from_pe_2, to_pe_2);
      // Instantiates other ring nodes and PEs...
    }
  \end{minted}
  \caption{Accelerator task instantiation in Vivado HLS.}
  \label{lst:ring-hls-inst}
\end{listing}

\subsection{Convenient Programming Interfaces}

\subsubsection{Communication Interface}

TAPA provides separate communication APIs for the producer side and
  the consumer side,
  which use \texttt{ostream} and \texttt{istream} as the interfaces,
  respectively.
The producer of a channel can test the fullness of the channel and append
  tokens to the channel (\texttt{write}) if the channel is not full.
The consumer of a channel can test the emptiness of the channel and remove
  tokens from the channel (destructive \texttt{read}),
  or duplicate the head of token without removing it (non-destructive read,
  a.k.a., \texttt{peek}),
  if the channel is not empty.
Read, peek, and write operations can be blocking or non-blocking.

A special token denoting end-of-transaction (EoT) is available to all
  channels.
A process can ``\texttt{close}'' a channel by writing an EoT token to it,
  and a process can ``\texttt{open}'' a channel by reading an EoT token from it.
A process can also test if a channel is closed,
  which is a non-destructive read operation to the channel (\texttt{eot}).
An EoT token does not contain any useful data.
This is designed deliberately to make it possible to break from a pipelined loop
  when an EoT is present,
  for example, in Line~\ref{lln:eot_tapa} of Listing~\ref{lst:ring-tapa-comm}.
Listing~\ref{lst:ring-tapa-comm} shows an example of how the
  communication interfaces are used in TAPA,
  which implements the same functionality as
  Listing~\ref{lst:ring-hls-comm},
  but with 55\% fewer lines due to the absence of the auxiliary \texttt{struct}
  for end-of-transaction token and the manually maintained input buffer that
  implements peek operations.

\begin{listing}[!t]
  \caption{Ring network node written in TAPA.}
  \label{lst:ring-tapa-comm}
  \inputminted[escapeinside=||,texcomments]{cpp}{code/communication.cpp}
\end{listing}

\subsubsection{Instantiation Interface}

A parent task can instantiate channels and tasks using the instantiation
  interface.
Channels are instantiated using \texttt{channel<type,capacity>}.
For example,
  \texttt{channel<Pkt,2>} instantiates a channel with capacity 2,
  and data tokens transmitted using this channel have type \texttt{Pkt}.
Tasks are instantiated using \texttt{task::invoke},
  with the first argument being the task function and the rest of arguments
  being the arguments to the task instance.
This is consistent with \texttt{std::invoke} in the C++ standard library.
Listing~\ref{lst:ring-tapa-inst} shows how channels and tasks are
  instantiated in TAPA.

\begin{listing}[!t]
  \caption{Accelerator task instantiation in TAPA.}
  \label{lst:ring-tapa-inst}
  \inputminted{cpp}{code/instantiation.cpp}
\end{listing}

\subsubsection{System Integration Interface}

TAPA uses a unified system integration interface to further reduce programmers'
  burden.
To offload a kernel to an FPGA accelerator,
  programmers only need to call the
  top-level task as a C++ function in the host code.
Since TAPA can extract metadata information, e.g., argument type,
  from the kernel code,
  TAPA will automatically synthesize proper OpenCL host API calls and emit an
  implementation of the top-level task C++ function that can set up the runtime
  environment properly.
As a user of TAPA,
  the programmer can use a single function invocation in the same source code to
  run software simulation, hardware simulation, and on-board execution,
  with the only difference of specifying proper kernel binaries.

\section{TAPA Framework Implementation}

\subsection{Software Simulation}
\label{sec:impl-sim}

\subsubsection*{State-of-the-Art Approach}

There are two state-of-the-art approaches to run software simulation
  for task-parallel applications:
  the sequential approach and the multi-thread approach.
A sequential simulator invokes tasks sequentially in the invocation
  order~\cite{man-vivado-hls}.
Sequential simulators are fast,
  but cannot correctly simulate the capacity of channels and applications with
  tasks communicating bidirectionally,
  as discussed in Section~\ref{sec:motivating-example}.
A multi-thread simulator invokes tasks in parallel by launching a thread for
  each task.
This enables the capacity of channels and bidirectional communication to be
  simulated correctly.
However,
  they may perform poorly due to the inefficient context switch handled by the
  operating system.
The FLASH simulator~\cite{fpga19-flash, tcad20-flash} proposed an alternative to
  the above,
  which uses HLS scheduling information to create an interleaving execution of
  all tasks.
Note that although FLASH is also single-threaded,
  it is different from a sequential simulator because it interleaves tasks via
  source-to-source transformation while a sequential simulator does not.
Compared with a sequential simulator,
  FLASH is on average 1.7{\texttimes} slower~\cite{tcad20-flash},
  due to additional scheduling information being taking into consideration for
  cycle-accurate modeling.
Besides,
  generating simulation executable becomes slower due to the need of the HLS
  scheduler output for cycle-accuracy,
  which is not needed for correctness verification.

In this section,
  we present an alternative approach to run software simulation on task-parallel
  applications.
Given that the inefficiency of multi-thread execution is mainly caused by the
  preemptive nature of operating system threads,
  we propose an approach that uses collaborative
  coroutines~\cite{toplas09-coroutine, man-boost-coroutine2} instead of
  preemptive threads for each task.
Note that fast and/or cycle-accurate debugging in
  general~\cite{trets20-fast-hls-debug} is out of
  the scope of this paper;
  we focus on the correctness and scalability issues for task-parallel programs.

\subsubsection*{Coroutine-Based Approach}
\label{sec:coroutines}

Routines in programming languages are the units of execution contexts, e.g.,
  functions in C/C++~\cite{book97-routine}.
Coroutines~\cite{commacm63-coroutine} are routines that execute collaboratively;
  more specifically, coroutines can be explicitly suspended and resumed.
A coroutine can invoke subroutines and suspend from and resume to any
  subroutine~\cite{man-boost-coroutine2}.
A context switch between coroutines takes only 26ns on modern
  CPUs~\cite{man-boost-coroutine2},
  while a preemptive thread context switch takes
  1.2{\textasciitilde}2.2{\textmu}s~\cite{web-pthread-context-switch},
  which is two orders of magnitude slower.

TAPA leverages coroutines to perform software simulation as follows.
When a task is instantiated,
  a coroutine is launched but suspended immediately.
Once all tasks are instantiated,
  the simulator starts to resume the suspended coroutines.
A resumed task will be suspended again if any input channel is accessed when
  empty or any output channel is accessed when full,
  which means that no progress can be made from this task.
A different task will then be selected and resumed by the simulator.
Moreover,
  the coroutines can be distributed in a thread pool.
The thread pool launches one thread per CPU core and can bind the thread to the
  corresponding core,
  which prevents the threads from preemption against each other.
This improves simulation parallelism without introducing high context switch
  overhead as in the multi-thread simulators.
We will show in Section~\ref{sec:exp-simulation} that the coroutine-based
  simulator outperforms the existing simulators by
  \csimtimespeedup{\texttimes} on average.
TAPA software simulator is implemented as a C++ library,
  which can be compiled by any compatible C++ compiler.

\subsection{RTL Code Generation}
\label{sec:impl-codegen}

\subsubsection*{State-of-the-Art Approach}

Current HLS tools treat the whole task-parallel program as a monolithic design,
  treat channels as global variables,
  and compile different instances of tasks as if they are completely unrelated.
This can lead to a significant amount of repeated work.
For example,
  the dataflow architecture generated by a stencil accelerator compiler,
  SODA~\cite{iccad18-soda, dac20-soda-cr},
  is highly modularized,
  and has many functionally identical modules.
However,
  both the Vivado HLS and Intel FPGA OpenCL backends generate RTL code for each
  module separately.
When the design scales out to hundreds of modules,
  RTL code generation can easily run for hours,
  taking even longer time than logic synthesis and implementation.
While we recognize that a programmer can manually generate RTL code for each
  task and glue them at RTL level,
  doing so defeats the purpose of using HLS for high productivity.
We also recognize that fast RTL code generation in general is an interesting
  problem,
  but we focus on the inefficiency exacerbated by task-parallel programs in this
  paper.

\subsubsection*{Modularized Approach}

Thanks to the hierarchical programming model,
  TAPA can keep the program hierarchy,
  recognize different instances of the same task,
  and compile each task only once.
As such,
  the total amount of time spent on RTL code generation is reduced.
Moreover,
  modularized compilation makes it possible to compile tasks in parallel,
  further reducing RTL code generation time on multi-core machines.
TAPA implements this by invoking the vendor tools in parallel for each task.
On average,
  TAPA reduces HLS compilation time by 4.9{\texttimes}
  (Section~\ref{sec:exp-codegen}).

\begin{figure}[!ht]
  \includegraphics[width=\linewidth]{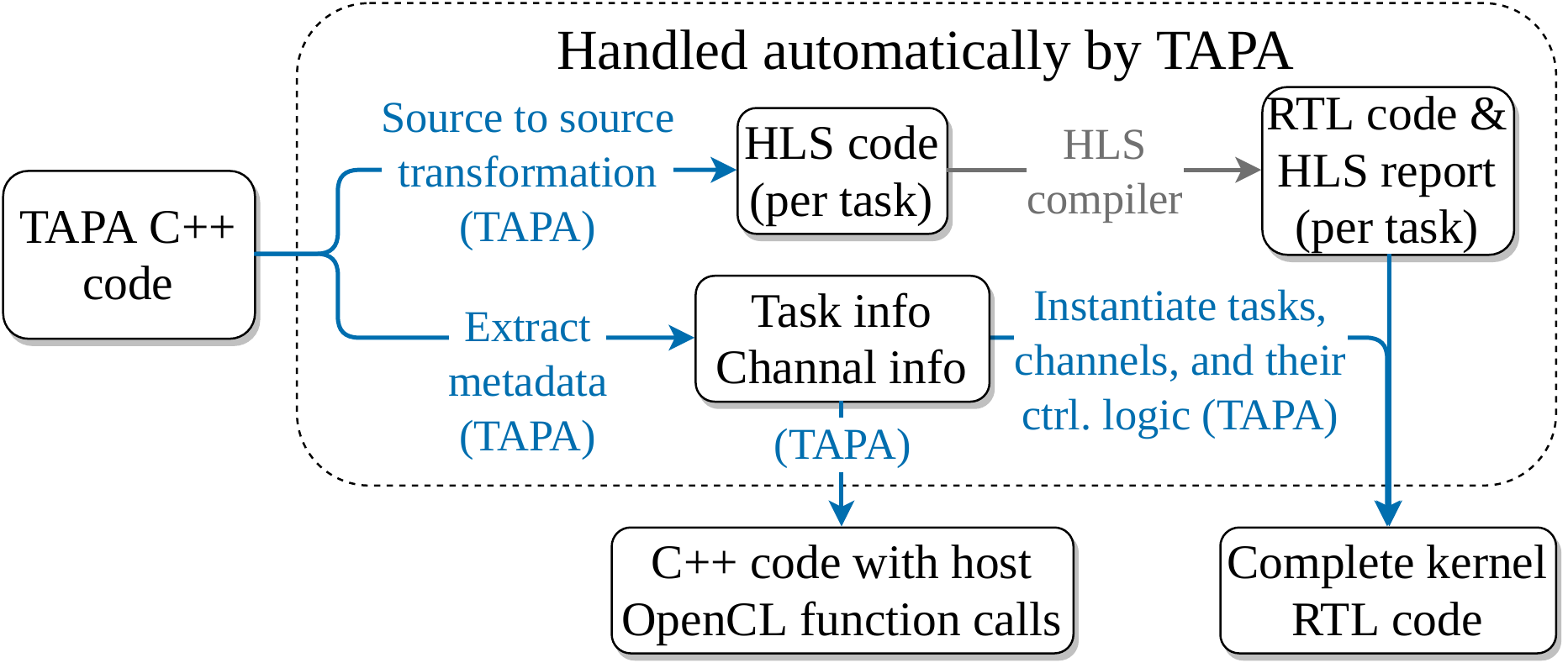}
  \caption{
    TAPA code generation.
    The host-kernel interface code is generated together with the kernel RTL
      code using metadata of the top-level task.
  }
  \label{fig:codegen}
\end{figure}

Fig.~\ref{fig:codegen} shows how RTL code is generated by TAPA,
  which is composed of four steps.
First,
  TAPA extracts the HLS code for each task and the metadata information of the
  whole design, including the communication topology among tasks,
  token types exchanged between tasks, and the capacity of each channel.
Source-to-source transformation is applied in this step to insert HLS pragmas
  where necessary (e.g., to generate proper RTL interfaces).
Then,
  the vendor HLS tool is used to generate RTL code and HLS report for each
  task.
While TAPA uses libraries to implement kernel APIs extensively,
  e.g., for \texttt{read}, \texttt{write}, and the end-of-transaction bit,
  not all APIs, e.g., peeking, can be implemented as libraries,
  due to the lack of support from the HLS scheduler.
To support peeking,
  TAPA adds a scalar argument to each \texttt{istream},
  and connect this port to the output of first-word-fall-through FIFO when
  the RTL code is assembled in the next step.

Using the metadata extracted in the first step,
  TAPA assembles the per-task RTL code to create the complete kernel.
In this step, for each parent task,
  TAPA instantiates the children tasks and channels,
  and generates a small state machine that controls start of the children tasks
  and termination of the parent task.
Finally,
  TAPA packages the assembled RTL code to a format that the vendor
  implementation tool can recognize (\texttt{xo} file for Vitis).

\section{Evaluation}

We prototype TAPA on Xilinx devices using Vivado HLS as the backend;
  support for Intel devices will be added later.
We compare the productivity of TAPA with two vendor tools that provide
  end-to-end high-level programming experience
  (including host-kernel communication):
  Xilinx Vitis 2019.2 suite and
  Intel FPGA SDK for OpenCL Pro Edition 19.4.
The experimental results are obtained on an
  Ubuntu 18.04 server with 2 Xeon Gold 6244 processors.

\subsection{Benchmarks}

Table~\ref{tab:benchmarks} summarizes the benchmarks used in this paper.
All implementations (Vivado HLS, Intel OpenCL, and TAPA) of each benchmark are
  written in such a way that tasks in each implementation have one-to-one
  correspondence,
  corresponding loops are scheduled with the same initiation interval (II),
  and each task performs the same computation.
This not only guarantees source codes to all tools are functionally equivalent,
  but also makes all tools generate consistent quality of results (QoR),
  which enables fair comparison of tool run time.
Note that we aim to compare the productivity of the HLS tools,
  not QoR (although we want to make sure there is no QoR degradation).
In particular,
  we were unable to guarantee that the generated RTL codes have exactly the same
  cycle-accurate behavior without having access to the HLS compiler's scheduling
  algorithm.
For example,
  the bucket sort network implemented in TAPA has a total latency of 3
  cycles while the Vivado HLS implementation has a total latency of 6.
This is inevitable because, using Vivado HLS,
  the manually maintained buffer forces an additional latency of 1 cycle at each
  network stage.
The shallower pipeline makes TAPA use 40\% fewer LUTs and 39\% fewer FFs for
  \texttt{network}.
For other benchmarks,
  TAPA uses 0.4\% fewer LUTs and 1\% fewer FFs on average.
This shows that the additional APIs provided by TAPA does not add resource
  overhead.

\begin{table}[!t]
  \centering
  \caption{
    Benchmarks used in this paper.
    Each task may be instantiated multiple times,
      so task instance count ({\#Inst.}) and channel count
      ({\#Chan.}) are greater than task count ({\#Task}).
  }
  \label{tab:benchmarks}
  {%
    \begin{tabular}{@{}ll@{}r@{}r@{}r@{}}
      \toprule
      Benchmark           & Application                                                       & \#Task~ & ~\#Inst.~ & ~\#Chan. \\
      \midrule
      \texttt{cannon}     & Cannon's algorithm~\cite{ics97-cannon}                            & 5       & 91        & 344      \\
      \texttt{cnn}        & VGG~\cite{iclr15-vgg} convolutional network~\cite{iccad18-polysa} & 14      & 209       & 366      \\
      \texttt{gaussian}   & Gaussian stencil filter~\cite{iccad18-soda}                       & 15      & 564       & 1602     \\
      \texttt{gcn}        & Graph convolutional network~\cite{iclr17-gcn}                     & 5       & 12        & 25       \\
      \texttt{gemm}       & General matrix multiplication~\cite{iccad18-polysa}               & 14      & 207       & 364      \\
      \texttt{network}    & Bucket sort w/ Omega network~\cite{toc75-omega-network}           & 3       & 14        & 32       \\
      \texttt{page\_rank} & PageRank citation ranking~\cite{tr98-pagerank}                    & 4       & 18        & 89       \\
      \bottomrule
    \end{tabular}
  }  %
\end{table}

\subsection{Lines of Kernel Code}

\begin{figure}[!t]
  \includegraphics[width=\linewidth]{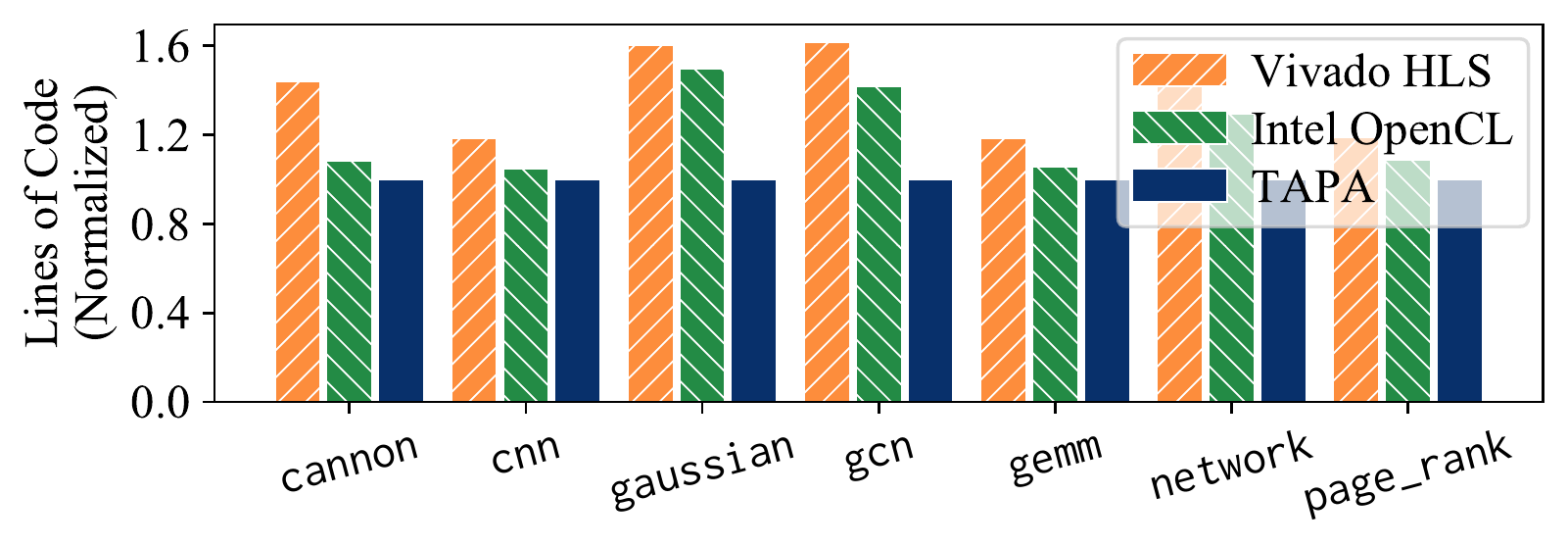}
  \caption{
    LoC comparison for kernel code. Lower is better.
  }
  \label{fig:kernel-loc}
\end{figure}

TAPA simplifies the kernel code in two aspects.
First,
  the TAPA communication interfaces simplify the code with the built-in support
  for peeking and transactions.
This not only simplifies the body of each task definition,
  but also removes the necessity for many \texttt{struct} definitions.
Second,
  the TAPA instantiation interfaces simplify the code by allowing tasks to be
  launched concisely.
Fig.~\ref{fig:kernel-loc} shows the lines of kernel code comparison of each
  benchmark.
On average, TAPA reduces the lines of kernel code by \kernellocreduced\%.
Note that only synthesizable kernel code is counted;
  code added for multi-thread software simulation
  is not counted for Vivado HLS.

\subsection{Lines of Host Code}

\begin{figure}[!t]
  \includegraphics[width=\linewidth]{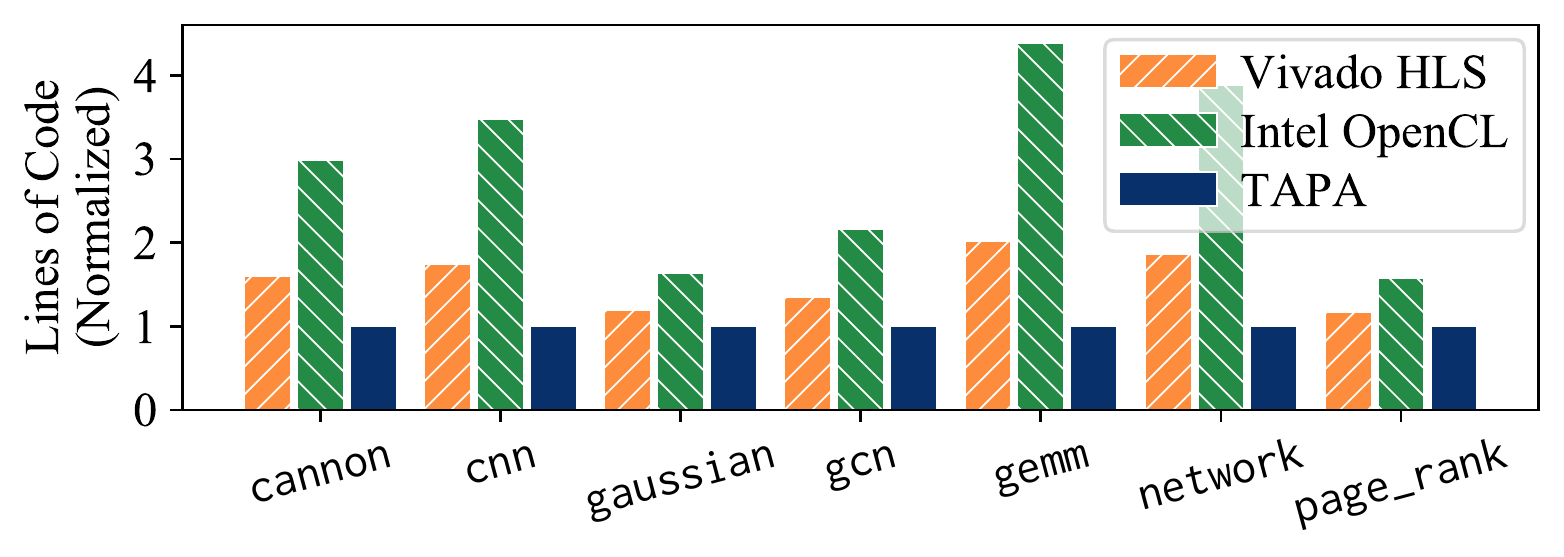}
  \caption{
    LoC comparison for host code. Lower is better.
  }
  \label{fig:host-loc}
\end{figure}

The host code used in the benchmarks contains a minimal test bench to verify the
  correctness of the kernel code.
TAPA system-integration API automatically interfaces with the OpenCL host APIs
  and relieves the programmer from writing repetitive
  code just to connect the kernel to a host program.
Table~\ref{fig:host-loc} shows the lines of host code comparison.
On average, the length of host code is reduced by \hostlocreduced\%.

\subsection{Software Simulation Time}
\label{sec:exp-simulation}

\begin{figure}[!t]
  \includegraphics[width=\linewidth]{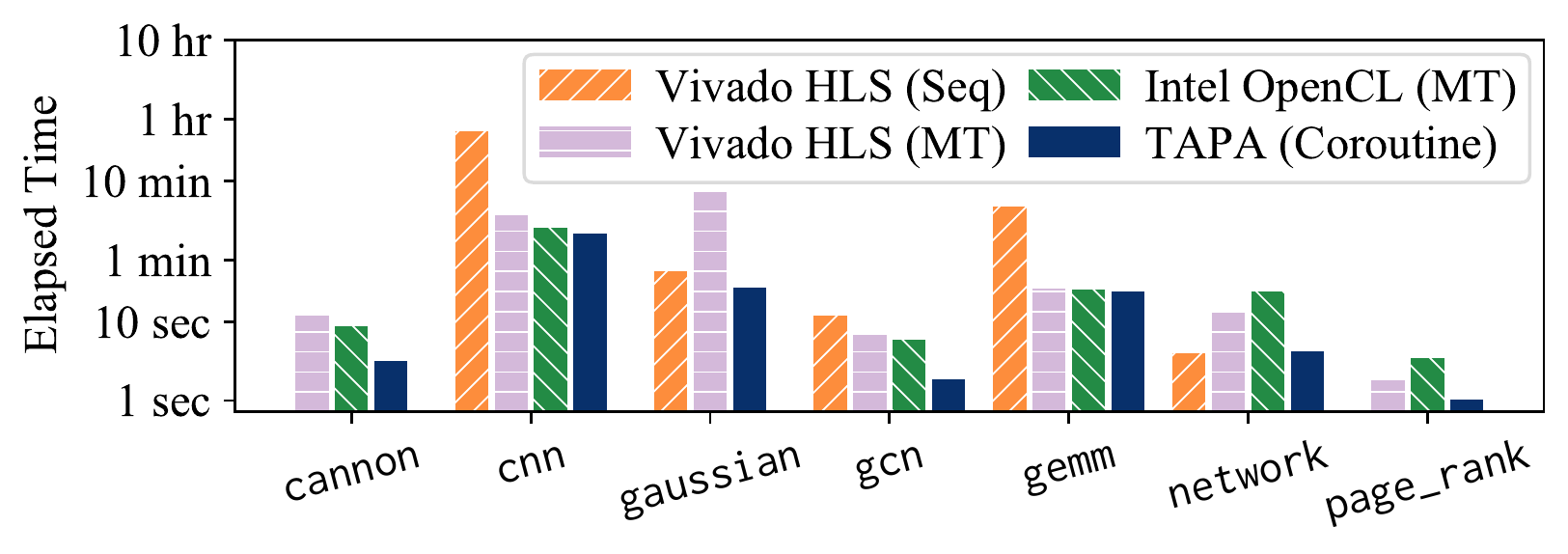}
  \caption{
    Simulation time in log scale.
    Lower is better.
    Sequential simulator fails to simulate \texttt{cannon} and
      \texttt{pagerank} correctly.
    Intel OpenCL multi-thread simulator cannot simulate \texttt{gaussian}
      due to its large number of task instances.
  }
  \label{fig:simulation}
\end{figure}

Fig.~\ref{fig:simulation} shows four simulators, that is,
  the sequential Vivado HLS simulator, the multi-thread Vivado HLS simulator,
  the multi-thread Intel OpenCL simulator,
  and the coroutine-based TAPA simulator.
Among the three simulators,
  the sequential simulator fails to correctly simulate benchmarks that
  require feedback data paths (\texttt{cannon} and \texttt{page\_rank}).
Due to the larger memory footprint required for storing the tokens transmitted
  between tasks and lack of parallelism,
  the sequential simulator is outperformed by the coroutine-based simulator in
  all but one of the benchmarks (\texttt{network}).
The two multi-thread simulators correctly simulate all benchmarks,
  except that Intel OpenCL cannot handle \texttt{gaussian} because its large
  number of task instances (564) exceeds the maximum allowed by the simulator
  (256).
However,
  the multi-thread simulators perform poorly on benchmarks that are
  communication-intensive
  (e.g., \texttt{network}) or have more tasks than the number of available
  threads (e.g., \texttt{gaussian}).
Although the coroutine-based TAPA simulator is not always the fastest simulator
  for all benchmarks, the worst-case slowdown is only 6\%,
  which is not significant in comparison with the multi-thread simulator,
  which can be 11{\texttimes} slower.
On average, TAPA is \csimtimespeedup{\texttimes} faster than other simulators.

\subsection{RTL Code Generation Time}
\label{sec:exp-codegen}

\begin{figure}[!t]
  \includegraphics[width=\linewidth]{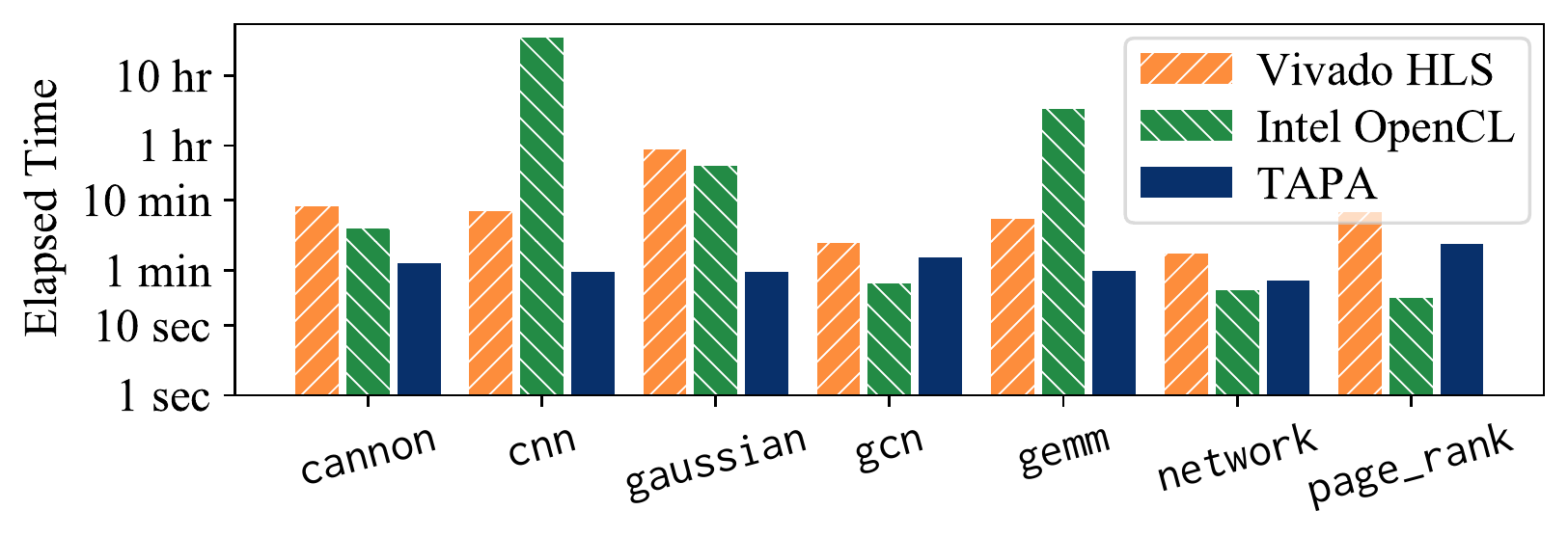}
  \caption{RTL code generation time in log scale. Lower is better.}
  \label{fig:codegen-time}
\end{figure}

Fig.~\ref{fig:codegen-time} shows the RTL code generation time comparison.
Thanks to the hierarchical programming model and modularized code generator,
  TAPA shortens the HLS compilation time by \hlstimespeedup{\texttimes} on
  average.
This is because
  \mycircled{1} TAPA runs HLS for each task only once even if it is instantiated
  many times, while Vivado HLS and Intel OpenCL run HLS for each task instance;
  \mycircled{2} TAPA runs HLS in parallel on multi-core machines.

\section{Related Work}
\label{sec:related-work}

Table~\ref{tab:related-work} on Page~\pageref{tab:related-work} shows a brief summary
  of the related HLS tools.
Section~\ref{sec:related-work-hls} presents more details about these tools.
Two domain-specific streaming frameworks are discussed in
  Section~\ref{sec:related-work-streaming}.
\textit{SystemC} and \textit{pthread} are two well-known alternative API
  paradigms that support task-parallel programs.
We will discuss and compare them with TAPA in
  Section~\ref{sec:related-work-alternative}.

\subsection{HLS Support for Task-Parallel Programs}
\label{sec:related-work-hls}

\textit{Intel HLS} supports two different inter-task communication
  interfaces: \texttt{pipe} and \texttt{stream}.
\texttt{pipe} implements a simple FIFO interface with \texttt{data},
  \texttt{valid}, and \texttt{ready} signals,
  while \texttt{stream} implements an
  Avalon-ST interface that supports transactions.
Tasks are instantiated using \texttt{launch} and \texttt{collect}.

\textit{Intel FPGA OpenCL} supports the simple FIFO interface via two
  sets of APIs,
  i.e., standard OpenCL \texttt{pipe} and Intel-specific \texttt{channel}.
Tasks are instantiated by defining OpenCL \texttt{\_\_kernel}s,
  which forces instances of the same task to be synthesized separately as
  different OpenCL kernels.

\textit{Vivado (Vitis) HLS} provides two different streaming interfaces:
  \texttt{ap\_fifo} and \texttt{axis}.
\texttt{ap\_fifo} generates the simple FIFO interface.
Tasks are instantiated by invoking the corresponding functions in a
  \texttt{dataflow} region (Listing~\ref{lst:ring-hls-inst}).
\texttt{axis} generates AXI-Stream interface with transaction support.
It requires the programmers to instantiate channels and
  tasks in a separate configuration file when running logic synthesis and
  implementation.
This allows different instances of the same task to be synthesized only once,
  but takes longer time to learn and implement compared with \texttt{ap\_fifo}.

\textit{Xilinx OpenCL} supports standard OpenCL \texttt{pipe},
  which generates AXI-Stream interfaces similar to Vivado HLS \texttt{axis},
  but \texttt{pipe} does not provide APIs to support transactions.

\textit{LegUp} supports the simple FIFO interface via \texttt{FIFO}.
Tasks are instantiated using \texttt{pthread} API
  (Section~\ref{sec:related-work-alternative}).

\textit{Merlin}~\cite{islped16-merlin} allows programmers to call the
  FPGA kernel as a C/C++ function and provides OpenMP-like simple pragmas with
  automated design space exploration based on machine learning.
To support task-parallel programs, Merlin leverages its backend vendor HLS
  tools' programming interfaces.

Their limitations are summarized in Table~\ref{tab:related-work} on
  Page~\pageref{tab:related-work}.
Note that a common limitation of HLS tools (including TAPA) is that they can not
  \textit{guarantee} the software description produces deterministic output
  sequences for task-parallel programs.
For instance,
  the emptiness test to an input channel is prone to breaking determinism,
  yet it is available to all HLS tools for performance and expressiveness
  reasons:
  merging two input channels round-robin using non-blocking reads would produce
  an output sequence determined by the relative arrival order of the input
  tokens.
An implication of non-determinism is we cannot assert that a program is
  deadlock-free just because its simulation succeeds.
This is different from deterministic programs, e.g.,
  Kahn process networks~\cite{ifip74-kpn},
  whose successful simulation generally implies deadlock-free on-board
  execution.
For applications that can be efficiently written without breaking determinism,
  e.g., streaming applications,
  there are dedicated frameworks developed specifically for them,
  which are discussed in the next section.

\subsection{Streaming Framework}
\label{sec:related-work-streaming}

\textit{ST-Accel}~\cite{fccm18-staccel} is a high-level programming platform
  that features highly efficient host-kernel
  communication interface exposed as a virtual file system (VFS).
It uses Vivado HLS as its backend for hardware generation.

\textit{Fleet}~\cite{asplos20-fleet} is a massively parallel streaming framework
  for FPGAs that features highly efficient memory interfaces for massive
  instances of parallel processing elements.
Programmers write Fleet programs in a domain-specific RTL language based on
  Chisel~\cite{dac12-chisel}.

TAPA aims to support more general task-parallel applications beyond streaming.

\subsection{Alternative APIs}
\label{sec:related-work-alternative}

\textit{SystemC} is a set of C++ classes and macros that provide detailed
  hardware modeling and event-driven simulation.
It supports both cycle-accurate and untimed simulation and many simulator
  implementations are available~\cite{dac17-systemc-sim, iscas14-systemc-sim}.
The official open-source SystemC simulator implementation uses coroutines
  without thread pooling.
Some HLS tools support a subset of untimed SystemC as the
  input~\cite{man-vivado-hls}.
SystemC supports task-parallel programs natively via the \texttt{SC\_MODULE}
  constructs and \texttt{tlm\_fifo} interfaces, which supports peeking.
While SystemC supports peeking FIFOs and coroutine-based simulation for
  task-parallel programs,
  it is limited by its special and verbose coding style.
Listing~\ref{lst:systemc} shows the example discussed in
  Section~\ref{sec:motivating-example} written in SystemC.
Compared with other C-like HLS languages,
  SystemC is more verbose and less
  productive due to its special language constructs:
  for TAPA code snippets shown in Listing~\ref{lst:ring-tapa-comm} and
  Listing~\ref{lst:ring-tapa-inst},
the equivalent SystemC kernel code would be 86\% longer.
On the host side,
  SystemC generates the main function in \texttt{sc\_main} by itself for
  simulation,
  and programmers need to spend time incorporating the SystemC test bench with
  other parts of their program.
This is not a problem if the whole system is defined by the kernel in SystemC,
  e.g., as in embedded systems,
  but in data center applications where the FPGA accelerator is only part of the
  system, this introduces non-trivial complication.

\begin{listing}[!t]
  \caption{
    SystemC TLM API example.
  }
  \label{lst:systemc}
  \inputminted{cpp}{code/systemc.cpp}
\end{listing}

\textit{Pthread} API is a set of widely used standard APIs that can be used to
  implement task-parallel programs using threads.
Pthread requires programmers to explicitly create and join threads,
  and each argument needs to be manually packed and passed.
Listing~\ref{lst:pthread} shows an example using the accelerator discussed in
  Section~\ref{sec:motivating-example}.
Compared with the \texttt{invoke} API used by TAPA,
  the pthread APIs require more effort to program:
  for TAPA code snippets shown in Listing~\ref{lst:ring-tapa-comm} and
  Listing~\ref{lst:ring-tapa-inst},
  equivalent pthread-based code would be 2.4{\texttimes} long.

\begin{listing}[!t]
  \caption{
    Pthread API example.
  }
  \label{lst:pthread}
  \inputminted{cpp}{code/pthread.cpp}
\end{listing}

In summary,
  while the API alternatives do exist in their own domains,
  they are more verbose and thus less productive compared with TAPA for
  task-parallel FPGA acceleration.

\section{Conclusion and Future Work}

In this paper,
  we present TAPA as an HLS C++ language extension to enhance the programming
  productivity of task-parallel programs on FPGAs.
TAPA has multiple advantages over state-of-the-art HLS tools: on average,
\mycircled{1}~its enhanced programming interface helps to reduce the lines of
  kernel code by \kernellocreduced\%,
\mycircled{2}~its unified system integration interface reduces the lines of host
  code by \hostlocreduced\%,
\mycircled{3}~its coroutine-based software simulator shortens the correctness
  verification development cycle by \csimtimespeedup{\texttimes},
\mycircled{4}~its modularized code generation approach shortens the QoR
  tuning development cycle by \hlstimespeedup{\texttimes}.
As a fully automated and open-source framework,
  TAPA aims to provide highly productive development experience for
  task-parallel programs using HLS.
For future work,
  we plan to extend our work to support dynamic
  tasks on FPGAs.

\section*{Acknowledgment}

The authors would like to thank the anonymous reviewers and our labmate,
  Linghao Song, for their valuable comments and helpful suggestions.
This work is partially supported by a Google Faculty Award,
  the NSF RTML program (CCF-1937599), NIH Brain Initiative (U01MH117079),
  the Xilinx Adaptive Compute Clusters (XACC) program,
  and CRISP, one of six JUMP centers.

\bibliographystyle{IEEEtran}
\bibliography{mendeley}

\end{document}